\documentclass[prb,aps,showpacs,twocolumn,unsortedaddress]{revtex4}
\usepackage{graphics,bm}
\usepackage{amssymb}
\usepackage{epsfig}
\usepackage{epsf}
\usepackage[usenames]{color}

\begin{document}

\title{Effects of non-adiabaticity on the voltage generated by a moving domain wall}

\author{R.A. Duine}
\email{R.A.Duine@uu.nl} \homepage{http://www.phys.uu.nl/~duine}

\affiliation{Institute for Theoretical Physics, Utrecht
University, Leuvenlaan 4, 3584 CE Utrecht, The Netherlands}
\date{\today}

\begin{abstract}
We determine the voltage generated by a field-driven domain wall,
taking into account non-adiabatic corrections to the motive force
induced by the time-dependent spin Berry phase. Both the diffusive
and ballistic transport regimes are considered. We find that that
the non-adiabatic corrections, together with the contributions due
to spin relaxation, determine the voltage for driving fields
smaller than the Walker breakdown limit.
\end{abstract}


\maketitle

\def\bx{{\bf x}}
\def\bk{{\bf k}}
\def\half{\frac{1}{2}}
\def\args{(\bx,t)}

\section{Introduction and summary of results} \label{sec:intro}
This paper develops the theory of non-adiabatic corrections to the
voltage generated by a moving domain wall. This aim is primarily
motivated by the fact that in certain situations, for example that
of a narrow wall, these corrections are important for the correct
description of current-driven domain-wall motion. We find that
non-adiabatic corrections play an important role in the reverse
process as well.

This introductory section is intended to be self-contained. The
more technical sections can be consulted for details of the
various calculations. We discuss current-driven magnetization
dynamics, and the reverse process, generation of current by a
time-dependent magnetization texture, in two separate subsections.
Our main results are presented and discussed at the end of this
section.

\subsection{Current-driven magnetization dynamics}
Let us consider a ferromagnetic metallic wire far below the Curie
temperature, characterized by a unit vector $\bm{\Omega} (\bx,t)$
in the direction of magnetization. Suppose we drive an electric
current through the wire, say in the $x$-direction, and that the
magnetization direction only varies in this direction. In the
adiabatic limit, that is, when the inverse Fermi wavenumber $k_F$
is much smaller than the typical length scale of
magnetization-direction variation, the electric transport current
is spin-polarized in the local direction of magnetization. As the
electrons traverse a non-collinear part of the ferromagnet, where
the magnetization direction changes from $\bm{\Omega} (x,t)$ to
$\bm{\Omega} (x+dx,t)$, the conduction electrons experience a
torque that changes their spin-polarization direction. This torque
is exerted by the magnetization. Conversely, there is a reaction
torque on the magnetization given by
\begin{equation}
\label{eq:sttsimpleexpression}
  \left. \frac{\partial \bm{\Omega} (x,t)}{\partial t} \right|_{\rm current}
  \propto \bm{\Omega} (x+dx,t) - \bm{\Omega} (x,t) \propto \frac{\partial \bm{\Omega}
  (x,t)}{\partial x}~,
\end{equation}
called a spin transfer torque.
\cite{slonczewski1996,berger1996,tsoi1998,myers1999,bazaliy1998,rossier2004,ralph2008}
This expression hinges on conservation of total spin. Roughly
speaking, the magnetization and conduction-electron spin precess
around each other while conserving their total spin angular
momentum. When there is electron-spin relaxation, in metals
primarily due to spin-orbit coupling and spin-flip scattering
events, there is an additional current-induced torque. In the
adiabatic limit this torque turns out to be of the form
$\bm{\Omega} (x,t) \times
\partial\bm{\Omega}/\partial x$, which is understood from the requirement that it should
be perpendicular both to the magnetization direction and to the
torque in Eq.~(\ref{eq:sttsimpleexpression}). In the adiabatic
limit the torques due to the current are sum of these two
contributions and given by
\begin{eqnarray}
\label{eq:sttsonly}
  \left. \frac{\partial \bm{\Omega} (x,t)}{\partial t} \right|_{\rm
  current} &=&\left( {\bf v}_s \cdot \bm{\nabla} \right) \bm{\Omega}
(\bx,t) \nonumber \\
&+&
 \beta_{\rm sr} \bm{\Omega} (\bx,t) \times \left( {\bf v}_s \cdot \bm{\nabla} \right) \bm{\Omega}
 (\bx,t)~,
\end{eqnarray}
with $\beta_{\rm sr}$ a dimensionless coefficient that
characterizes the degree to which spin is not conserved in the
spin-transfer process, \cite{zhang2004,barnes2005,
tserkovnyak2006,kohno2006,piechon2006,duine2007} and the velocity
${\bf v}_{\rm s} = - {\mathcal P} {\bf j}/|e| \rho_s $ is
proportional to the current ${\bf j}$ (with ${\mathcal P}$ the
current spin polarization, $\rho_s $ the density difference of
majority and minority spin electrons, and $-|e|$ the electron
charge). The two terms on the right-hand side of the above
equation are properly called the reactive and dissipative
adiabatic spin transfer torques, \cite{duine2007,tserkovnyak2008}
respectively, although they are also referred to as adiabatic and
non-adiabatic for reasons that will become clear shortly.

\begin{center}
\begin{figure}
\epsfig{figure=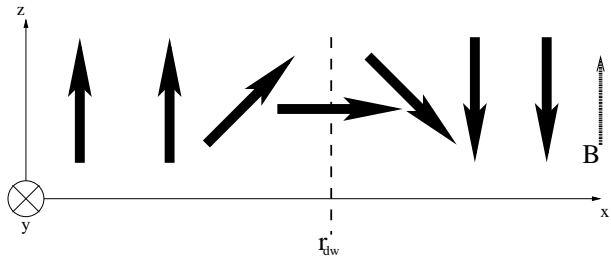, width=8cm}
 \caption{Illustration of a domain wall at position $r_{\rm dw} (t)$ in a magnetic field $B$ pointing in the $z$-direction.
 The chirality $\phi_0 (t)$ is the angle of the spin at the position of the domain wall with the $x$-$y$-plane.}
 \label{fig:dw}
\end{figure}
\end{center}

In a series of papers, \cite{tatara2004,tatara2007,tatara2008}
Tatara, Kohno, Shibata, and co-authors, have also considered
non-adiabatic corrections to Eq.~(\ref{eq:sttsonly}),
\cite{footnote1} that quite generally contribute a non-local term
to Eq.~(\ref{eq:sttsonly}) so that it becomes
\begin{eqnarray}
\label{eq:standmt}
  &&\left. \frac{\partial \bm{\Omega} (x,t)}{\partial t} \right|_{\rm
  current} =\left( {\bf v}_s\!\cdot\!\bm{\nabla} \right) \bm{\Omega}
(\bx,t) \nonumber \\ &&+
 \beta_{\rm sr} \bm{\Omega} (\bx,t)\!\times\!\left( {\bf v}_s\!\cdot\!\bm{\nabla} \right) \bm{\Omega}
 (\bx,t)  + \Gamma_{\rm na} \left[\bm{\Omega}\right],
\end{eqnarray}
with $\Gamma_{\rm na} [\bm{\Omega}]$ a functional that can in
principle be evaluated in certain limits. (See also
Refs.~[\onlinecite{waintal2004}],~[\onlinecite{thorwart2008}],~and~[\onlinecite{nguyen2007}]
for a treatment of non-adiabatic corrections.) In particular,
Tatara {\it et al.} \cite{tatara2004,tatara2007,tatara2008}
considered the effect of momentum transfer, which corresponds
physically to electrons scattering off the magnetization texture.
The evaluation of the non-local torque $\Gamma[\bm{\Omega}]$ that
corresponds to this process is quite complicated for a general
magnetization texture. Motivated by ongoing
experimental\cite{grollier2003,tsoi2003,yamaguchi2004,
klaui2005,beach2006,hayashi2007,yamanouchi2004,yamanouchi2006} and
theoretical\cite{tatara2004,tatara2007,tatara2008,zhang2004,waintal2004,barnes2005,thiaville2005,rebei2005,ohe2006,xiao2006,tserkovnyak2006,
kohno2006,piechon2006,duine2006,nguyen2007} research on
current-driven motion of domain walls,  Tatara {\it et al.}
\cite{tatara2004,tatara2007,tatara2008}  evaluated
Eq.~(\ref{eq:standmt}) within a simple variational description of
the domain wall. Namely, they showed that for a rigid domain wall
Eq.~(\ref{eq:standmt}) yields
\begin{eqnarray}
\label{eq:cidwvariational} \left. \frac{d\phi_0 (t)}{d t}
\right|_{\rm current}
 &=& \left( \beta_{\rm sr} + \beta_{\rm na} \right) \frac{v_s}{\lambda}~;
 \nonumber \\
 \left. \frac{d r_{\rm dw} (t)}{d t} \right|_{\rm current}
 &=& v_s~,
\end{eqnarray}
where the current is taken in the direction of the domain wall,
i.e., in the direction of varying magnetization. In the above
expression the dynamical variational parameters $r_{\rm dw} (t)$
and $\phi_0 (t)$, denote the position of the domain wall, and its
chirality, respectively. For the case of an easy-plane
ferromagnet, for example, $\phi_0 (t)$ is the canting angle with
which the magnetization tilts out of the easy plane at the
domain-wall position. Furthermore, $\lambda$ is the width of the
domain wall, determined by the competition between exchange and
anisotropy. See Fig.~\ref{fig:dw} for an illustration.

In the clean limit, the coefficient $\beta_{\rm na}$ is found as
\cite{tatara2007}
\begin{equation}
\label{eq:betamt}
 \beta_{\rm na} = \frac{\lambda}{{\mathcal P}A} \frac{|e|^2}{\hbar} N_e
 \rho_{\rm dw}~,
\end{equation}
with $A$ the cross-section of the magnetic wire perpendicular to
the domain wall and current direction, $N_e$ the number of
electrons, and $\rho_{\rm dw}$ the contribution to the resistivity
due to the domain wall.

Interestingly, the momentum transfer process, described by a
complicated non-local term in Eq.~(\ref{eq:standmt}) for the full
magnetization direction dynamics, yields a simple renormalization
of $\beta_{\rm sr}$ at the level of the variational description in
Eq.~(\ref{eq:cidwvariational}).\cite{tatara2008} (This is the
reason for referring to the term proportional to $\beta_{\rm sr} $
in Eq.~(\ref{eq:standmt}) as ``non-adiabatic", which is strictly
speaking incorrect.) The main result of this paper is that the
same renormalization occurs for the case of charge-current
generation by a moving domain wall.

We end this subsection by mentioning that, over the last few
years, Eqs.~(\ref{eq:standmt})~and~(\ref{eq:cidwvariational}) have
been actively debated. In particular, the ratio of the coefficient
$\beta_{\rm sr}$ to the so-called Gilbert damping constant
$\alpha_G$ that governs magnetization relaxation has been subject
of interest. Although it is now generally accepted that this ratio
is generally not equal to one, as indicated by microscopic
theories \cite{kohno2006,duine2007} and recent experiments,
\cite{heyne2008} characterizing and optimizing the various
processes that lead to current-driven domain-wall motion will most
likely remain an active topic of research in the near future.

\subsection{Current generated by a time-dependent magnetization texture}
The reverse process of current-driven magnetization dynamics is
the generation of current and voltage by a time-dependent
magnetization texture. The expression for the charge current is
given by \cite{stern1992,barnes2007,duine2008,tserkovnyak2008b}
\begin{eqnarray}
\label{eq:resultelectriccurrent} &&  j_\alpha =
-\frac{\hbar}{2|e|V}\left(
  \sigma_\uparrow\!-\!\sigma_\downarrow \right) \nonumber \\
  &&\times \left[ \frac{\partial}{\partial t} \int d\bx \tilde
A_{\alpha'}(\bm{\Omega}(\bx,t)) \nabla_\alpha
  \Omega_{\alpha'} (\bx,t)
 \right.\nonumber \\
&&  \left . +   \beta_{\rm sr}  \int d\bx
   \frac{\partial \bm{\Omega} (\bx,t)}{\partial t} \cdot
   \nabla_\alpha \bm{\Omega} (\bx,t) + {\mathcal E}_{\rm na}
   [\bm{\Omega}]
  \right]~,
\end{eqnarray}
with $V$ the volume of the system and $\sigma_\uparrow$ and
$\sigma_\downarrow$ the respective conductivities of the majority
and minority spin electrons. In this paper, a summation over
repeated indices $\alpha,\alpha',\alpha'' \in \{ x,y,z\}$ is
implied.

The three terms in the above equation correspond, roughly
speaking, to the three terms on the right-hand side of
Eq.~(\ref{eq:standmt}), respectively. That is, the first term is
adiabatic and, loosely speaking, due to conservation of spin. It
is given in terms of a vector potential $\tilde A_{\alpha}
(\bm{\Omega})$ of a magnetic monopole in spin space (not to be
confused with the electromagnetic vector potential ${\bf A}$ that
we will introduce later on) that obeys
$\epsilon_{\alpha,\alpha',\alpha''}
\partial \tilde A_{\alpha'}/\partial \Omega_{\alpha''}=\Omega_\alpha$ and is well-known
from the path-integral formulation for spin systems.
\cite{auerbachbook} This term corresponds to the time-derivative
of the flux of a monopole magnetic field (in spin space) enclosed
by the path $\bm{\Omega} (\bx,t)$ on the unit sphere, and is the
motive force induced by the time-dependent spin Berry phase. It
was first discussed by Stern, \cite{stern1992} and later by Barnes
and Maekawa in the context of Faraday's law in a ferromagnetic
metal. \cite{barnes2007}

Before discussing the remaining two contributions to the generated
current, we give, for completeness and future reference, first an
expression for the voltage drop $\Delta V$ in the
$\alpha$-direction across a wire of length $L$ with
cross-sectional area $A$. Using that $\Delta V = j_\alpha
L/(\sigma_\uparrow + \sigma_\downarrow)$, we find
\begin{eqnarray}
\label{eq:finalresultvoltage}  &&  \Delta V  =
-\frac{\hbar {\mathcal P}}{2|e|A} \nonumber \\
  &&\times \left\{   \int d\bx
  \bm{\Omega} (\bx,t) \cdot \left[ \frac{\partial \bm{\Omega} (\bx,t)}{\partial t}
  \times \nabla_\alpha \bm{\Omega} (\bx,t) \right]  \right.
\nonumber \\
&&  \left . +   \beta_{\rm sr}  \int d\bx
   \frac{\partial \bm{\Omega} (\bx,t)}{\partial t} \cdot
   \nabla_\alpha \bm{\Omega} (\bx,t) + {\mathcal E}_{\rm na}
   [\bm{\Omega}]
  \right\}~,
\end{eqnarray}
where we used the properties of the vector potential $\tilde {\bf
A} (\bm{\Omega})$ to work out the first term in
Eq.~(\ref{eq:resultelectriccurrent}), and that ${\mathcal P}
\equiv
(\sigma_\uparrow-\sigma_\downarrow)/(\sigma_\uparrow+\sigma_\downarrow)$.
Note that in deriving the expression for the voltage we have
assumed that the total conductivity is given by
$\sigma_\uparrow+\sigma_\downarrow$, and have therefore neglected
other contributions, e.g., due to the presence of a domain wall
and/or anisotropic magnetoresistance. We will come back to the
point in the following sections.

The second term in
Eqs.~(\ref{eq:resultelectriccurrent},\ref{eq:finalresultvoltage})
was first derived in Ref.~[\onlinecite{duine2008}] using
response-function techniques. It corresponds to an adiabatic
correction, due to spin-orbit coupling and spin-flip scattering,
to the Berry-phase-induced motive force (which is also adiabatic).
The same correction was found by Tserkovnyak and Mecklenburg
\cite{tserkovnyak2008b} using Onsager reciprocity. We mention also
the work by Saslow \cite{saslow2007} who considered the generation
of electric current by time-dependent magnetization textures in
the absence of conservation of spin, within the framework of
irreversible thermodynamics.

The last contribution to
Eqs.~(\ref{eq:resultelectriccurrent},\ref{eq:finalresultvoltage}),
proportional to ${\mathcal E}_{\rm na} [\bm{\Omega}]$, corresponds
formally to all terms beyond linear order in spatial gradients of
the magnetization direction. Although, to the best of our
knowledge, it has not been considered in great detail, it could be
calculated for example order by order in a gradient expansion. The
main result of this paper is that, when
Eqs.~(\ref{eq:resultelectriccurrent},\ref{eq:finalresultvoltage})
are evaluated within the same description that yields
Eq.~(\ref{eq:cidwvariational}) from Eq.~(\ref{eq:standmt}), one
finds that \cite{footnote2}
\begin{equation}
\label{eq:resultcurrentafteransatz2}
 \Delta V  = -\frac{\hbar}{|e|} \left(\frac{\sigma_\uparrow-\sigma_\downarrow}{\sigma_\uparrow+\sigma_\downarrow}
  \right)\left[\dot \phi_0 (t)-
  \frac{\left(\beta_{\rm sr}+\beta_{\rm na}\right) \dot r_{\rm dw} (t)}{\lambda}
  \right]~.
\end{equation}
Hence, taking into account non-adiabatic corrections to the
generated current again amounts to the renormalization $\beta_{\rm
sr} \to \beta_{\rm sr}+\beta_{\rm na}$ at the level of the
effective (variational) model. To arrive at this result it turns
out to be sufficient to assume Eq.~(\ref{eq:cidwvariational}),
from which the above result can be straightforwardly derived, as
we show in the following section.

\begin{center}
\begin{figure}
\epsfig{figure=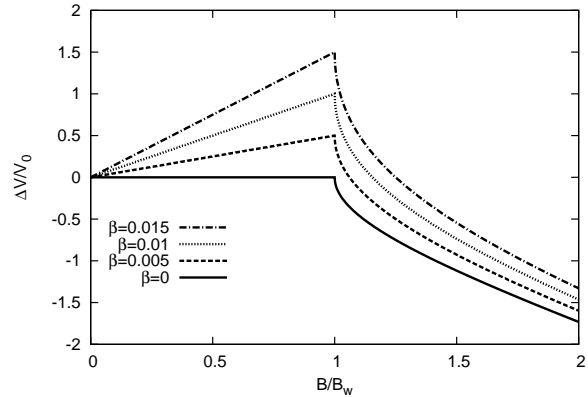, width=8cm}
 \caption{Voltage
  induced by a moving domain wall for Gilbert damping $\alpha_G=0.01$
  and various values of $\beta \equiv \beta_{\rm sr} + \beta_{\rm na}$. The voltage is normalized
  to $V_0 = {\mathcal P} g \mu_B B_w/|e|$.
  The magnetic field $B$ is in unit of the Walker breakdown field $B_w$.
  For typical experiments we have that $V_0 \sim 0.5$ $\mu$Volt.}
 \label{fig:v_b}
\end{figure}
\end{center}

Next, we give the result for the evaluation of
Eq.~(\ref{eq:resultcurrentafteransatz2}) for the case of a
field-driven domain wall with an easy-plane and a hard axis. In
this case, the wall moves with constant velocity below the
so-called Walker breakdown field $B_w$. \cite{schryer1974} This
field is proportional to the hard axis anisotropy energy, and to
the Gilbert damping constant $\alpha_G$. The domain-wall
precession angle is time-independent for fields $B$ smaller than
$B_w$. Above this field the domain-wall chirality becomes
time-dependent and the domain wall undergoes oscillatory motion.
The result in Eq.~(\ref{eq:resultcurrentafteransatz2}) leads to
\cite{duine2008}
\begin{eqnarray}
\label{eq:finalresultwithinapprox} && \frac{\Delta V}{V_0} =
  \frac{\left(\beta_{\rm sr}+\beta_{\rm na}\right)}{\alpha_{G}} \frac{B}{B_w} \nonumber \\
 &&  -\left( \frac{1+ \frac{\left(\beta_{\rm sr}+\beta_{\rm na}\right)}{\alpha_{\rm G}}}{1+\alpha_{\rm G}^2} \right) {\rm Re} \left[
 \sqrt{\left( \frac{B}{B_w}\right)^2\!\!-1}\right]~,
\end{eqnarray}
with $V_0 = {\mathcal P} g \mu_B B_w/|e|$, in terms of the
gyromagnetic ratio $g$ and the Bohr magneton $\mu_B$. For typical
experiments\cite{beach2005} ($B_w \sim 100$ Oe) we have that $V_0
\sim 0.5$ $\mu$Volt. Note that $V_0$ is roughly the Walker
breakdown field converted to a voltage. In Fig.~\ref{fig:v_b} we
plot this result for various values of $(\beta_{\rm sr} +
\beta_{\rm na})/\alpha_G$. From this figure, and also from the
expression in Eq.~(\ref{eq:finalresultwithinapprox}), it is
obvious that below Walker breakdown the induced voltage is
completely determined by spin-orbit coupling and spin-flip
scattering, and the by the non-adiabatic corrections.

In Sec.~\ref{sec:diffusive} we derive our main result in
Eq.~(\ref{eq:resultcurrentafteransatz2}). In
Sec.~\ref{sec:ballistic} we investigate how the two contributions
to the generated voltage, proportional to $dr_{\rm dw}/dt$ and
$d\phi_0/dt$, arise in a ballistic model for electron transport.
We end in Sec.~\ref{sec:disc} with a short discussion and outlook.
Before we turn to the more technical part of this paper, let us
end this section with a brief description of other work on the
generation of charge current by a time-dependent magnetization.
Already in 1986, Berger \cite{berger1986} discussed the generation
of current by moving domain walls in terms of an analogue of the
Josephson effect. The motive force due to the time-dependent spin
Berry phase was first pointed out by Stern \cite{stern1992} in the
context of a mesoscopic ring with a Zeeman magnetic field, and
later considered in the context of domain walls by Barnes and
Maekawa \cite{barnes2007}. Both these works did not consider the
effects of spin-orbit coupling and spin-flip scattering in the
adiabatic limit. This was done first in
Ref.~[\onlinecite{duine2008}], by Saslow, \cite{saslow2007} and by
Tserkovnyak and Mecklenburg. \cite{tserkovnyak2008b} The limit of
strong Rashba spin-orbit coupling was considered by Ohe {\it et
al.} \cite{ohe2007} Yang {\it et al.}\cite{yang2007} evaluated
Eq.~(\ref{eq:finalresultvoltage}) with $\beta_{\rm sr}=0$ and
without non-adiabatic corrections using results of micromagnetic
simulations for straight and vortex domain walls. Stamenova {\it
et al.} \cite{stamenova2008} performed a detailed numerical
analysis of the Berry-phase-induced motive force, also without
considering spin-orbit coupling or spin-flip scattering.

Although there are, to the best of our knowledge, at present no
published experimental results on the generation of current and
voltage by a moving domain wall, such results would present an
important step forward in completing our understanding knowledge
of the interplay between electric current and magnetization
dynamics. We hope and expect that the above-mentioned theoretical
work will be confronted with experimental results in the near
future.

\section{Diffusive transport regime} \label{sec:diffusive} In the
diffusive transport regime, we assume that the ferromagnetic metal
is characterized by conductivities $\sigma_\uparrow$ and
$\sigma_\downarrow$, for the majority and minority spin bands,
respectively. Furthermore, we assume that the velocity $v_s$ is
given by the linear-response expression
\begin{equation}
\label{eq:vsinlinearresponse}
  v_s = -\frac{ \left( \sigma_\uparrow-\sigma_\downarrow \right)E}{|e|
  \rho_s}~,
\end{equation}
where the electric field $E$ is taken in the domain-wall
direction. The easiest way to proceed is by writing down an action
${\mathcal A} [r_{\rm dw}, \phi_0]$ that, upon variation,
reproduces the equation of motion for the variational parameters
in Eq.~(\ref{eq:cidwvariational}). Using the short-hand notation
\begin{equation}
\beta \equiv \left( \beta_{\rm sr} + \beta_{\rm na} \right)~,
\end{equation}
we have in imaginary time $\tau=it$ that
\begin{eqnarray}
\label{eq:actionforrandphi}
  {\mathcal A} [r_{\rm dw},\phi_0] &=& \int_0^{\hbar/k_B T}  d\tau
  N \left[ i \hbar\frac{r_{\rm dw} (\tau)}{\lambda} \frac{d\phi_0
  (\tau)}{d\tau}\right.
  \nonumber \\
  &&\left.+\hbar \frac{v_s}{\lambda} \phi_0 (\tau) - \hbar \beta v_s \frac{ r_{\rm dw}
  (\tau)}{\lambda^2}\right]~,
\end{eqnarray}
with $k_B T$ the thermal energy. Here, $N=\lambda A \rho_s$ is the
number of electron spins in the domain wall. In principle, the
action for the domain wall contains potential-energy terms due to
anisotropy and external magnetic field. They are, however, not
important in describing the coupling of the current to the domain
wall and are omitted in the above expression [as well as in
Eq.~(\ref{eq:cidwvariational})].

Although the above action can in principle be derived
microscopically within a given approximation scheme, this is not
needed here. All we need to extract the contributions to the
current due to the moving domain wall is the fact that there
exists a response function $\Pi (\bx,\bx';\tau-\tau')$, such that
\begin{equation}
\label{eq:vsrespfct}
  v_s = \int_0^{\hbar/k_B T}\!d\tau' \int\!d\bx \int\!d\bx'\Pi (\bx,\bx';\tau-\tau') A (\bx',\tau')~,
\end{equation}
reduces to Eq.~(\ref{eq:vsinlinearresponse}) for a vector
potential
\begin{equation}
\label{eq:vecpot}
  A (\bx,\tau) = -\frac{cE}{\omega_p} e^{-i\omega_p \tau}~,
\end{equation}
when we take the zero-frequency limit $\omega_p \to 0$. Here, $c$
is the velocity of light. This requirement determines the
low-frequency behavior of the response function $\tilde \Pi (i
\omega_n) $, defined by,
\begin{eqnarray}
\label{eq:respfctfreq}
 && \frac{1}{\hbar/k_BT} \sum_n \tilde \Pi (i\omega_n)
  e^{-i \omega_n (\tau-\tau')} \nonumber \\
  && =\int\!d\bx\int\!d\bx' \Pi (\bx,\bx';\tau-\tau')~,
\end{eqnarray}
as
\begin{equation}
\label{eq:lowfreqresfct}
 \tilde  \Pi (i \omega_n) = \frac{\left(\sigma_\uparrow-\sigma_\downarrow
  \right)\omega_n}{c|e|\rho_s}~,
\end{equation}
with $\omega_n = 2\pi n k_B T/\hbar$ the bosonic Matsubara
frequencies. At this point we note that the non-adiabaticity is,
although somewhat hidden in the formalism, incorporated by
allowing the response function $\Pi (\bx,\bx';\tau-\tau')$ to
depend arbitrarily on the spatial coordinates $\bx, \bx'$.

The next ingredient we need is that quite generally the
expectation value of the electric current $j$ is given by a
functional derivative of the effective action via
\begin{equation}
\label{eq:currentasderivativeeffaction}
  j = c \frac{\delta {\mathcal A} [r_{\rm dw},\phi_0]}{\delta A
  (\bx,\tau)}~.
\end{equation}
This yields in first instance for the current
\begin{eqnarray}
\label{eq:currentfirst}
  j &=& \frac{c}{L} \int_0^{\hbar/k_B T}\!d\tau' \int\!d\bx
 \int\!d\bx' \rho_s \hbar\nonumber \\
 && \times \left\{ \left[ \phi_0 (\tau') - \beta
\frac{r_{\rm dw} (\tau') }{\lambda }\right]
  \Pi (\bx,\bx';\tau'-\tau) \right\}~.
\end{eqnarray}
Using now the result in Eq.~(\ref{eq:lowfreqresfct}) we find in
the low-frequency limit that
\begin{eqnarray}
\label{eq:intermediateresultvarpars}
  j = - i \frac{\hbar}{|e|L} \left(\sigma_\uparrow-\sigma_\downarrow \right) \left[ \frac{d\phi(\tau)}{d\tau}- \beta \frac{dr_{\rm dw} (\tau)}{\lambda
  d\tau}\right]~,
\end{eqnarray}
which, after a Wick rotation $\tau=it$ and realizing that $\Delta
V = jL/(\sigma_\uparrow +\sigma_\downarrow)$, leads to
Eq.~(\ref{eq:resultcurrentafteransatz2}).

This result is rather general, in the sense that it does not
depend on specific values of $\alpha_G$ and $\beta_{\rm sr} +
\beta_{\rm na}$ or the microscopic mechanisms contributing to
these coefficients. The only input is that
Eq.~(\ref{eq:cidwvariational}), together with Ohm's law in
Eq.~(\ref{eq:vsinlinearresponse}), holds. Moreover, the result is
applicable both within the $s-d$ and the Stoner model for
ferromagnetism. Finally, it is important to note that in
calculating the voltage $\Delta V$, we have neglected the
contribution to the conductivity due to the presence of the domain
wall. In the ballistic limit, to be discussed in the next section,
this contribution is accounted for rather straightforwardly.

\section{Ballistic limit} \label{sec:ballistic} In this section we
show how the two contributions to the voltage in
Eq.~(\ref{eq:resultcurrentafteransatz2}), proportional to
$d\phi_0/dt$ and $dr_{\rm dw}/dt$, respectively, arise in the
ballistic limit where the scattering theory of electronic
transport \cite{buttiker1985} is applicable. This is instructive
as the discussion of the generation of spin and charge currents in
mesoscopic systems, called spin pumping and charge pumping,
respectively, is usually done within this framework. Applications
involve quantum dots,
\cite{bruder1994,brouwer1998,switkes1999,watson2003} and
single-domain ferromagnets.
\cite{tserkovnyak2002,tserkovnyak2002b,mizukami2001,costache2006,xiao2008}

The starting point is the expression due to B\"uttiker {\it et
al.} \cite{buttiker1994,tserkovnyak2002b} that gives the current
in terms of derivatives of the scattering matrix
\begin{equation}
\label{eq:currentviascatmatrix}
  I_\nu= -\frac{|e|}{4\pi i} \sum_{\gamma,\nu'} {\rm Tr} \left[
  {\bf s}_{\nu\nu'}^\dagger \cdot \frac{\partial {\bf s}_{\nu\nu'}}{\partial
  X_\gamma}- \frac{\partial {\bf s}_{\nu\nu'}^\dagger}{\partial
  X_\gamma} \cdot  {\bf s}_{\nu\nu'}
  \right] \frac{d X_\gamma}{dt}~,
\end{equation}
to lowest order in the time derivatives. We note at this point
that, although the above expression is first-order in
time-derivatives, it contains all non-adiabatic corrections
\cite{footnote1} because it depends on the scattering matrix which
in turn depends on the full magnetization texture $\bm{\Omega}
(x)$ and not only its first-order spatial derivative
$\partial\bm{\Omega} (x)/\partial x$. (Note, however, that in this
section we ignore the contribution due to spin relaxation to the
induced current and voltage.) In this expression, $I_\nu$ is the
current \cite{footnote3} into reservoir $\nu$, and the sum $\nu'
\in \{1,2\}$ runs over the left and right reservoir (lead),
denoted by $1$ and $2$, respectively. The $X_\gamma$ label the
parameters that vary in time and change the scattering matrix, and
the sum over $\gamma$ is over all such parameters.

The scattering matrix
\begin{eqnarray}
 {\bf s} =\left( \begin{array}{cc}
      {\bf r}_{11} & {\bf t}_{12} \\
      {\bf t}_{21} & {\bf r}_{22}
    \end{array} \right)~,
\end{eqnarray}
is given in terms of the reflection amplitudes ${\bf r}_{\nu\nu}$,
and transmission amplitudes ${\bf t}_{\nu\nu'}$ that describe
transmission from reservoir $\nu'$ to $\nu$. These quantities have
matrix structure in the space of conduction channels of the leads.
In Eq.~(\ref{eq:currentviascatmatrix}) they are to be evaluated at
the Fermi energy  $\epsilon_F$ of the leads.

Consider now specifically the domain-wall configuration shown in
Fig.~\ref{fig:dw}, with $\phi_0=0$ and with the domain-wall
located at the origin. We parameterize this domain-wall
magnetization-direction texture by $\bm{\Omega}_{\rm dw} (x)=(\sin
\theta_{\rm dw} (x), 0, \cos \theta_{\rm dw} (x))$. Within the
simplest two-band mean-field model for the ferromagnetism (which
could be extended to more complicated and realistic situations,
see Ref.~[\onlinecite{vanhoof1999}]), the wave function $\psi
(x)\equiv(\psi_\uparrow (x), \psi_\downarrow (x))$ of the
electrons with energy $\epsilon$ moving in the presence of this
texture, obeys the time-independent Schr\"odinger equation
\begin{equation}
\label{eq:se}
   \left[ -\frac{\hbar^2}{2m} \frac{d^2}{dx^2} - \frac{\Delta}{2} \bm{\Omega}_{\rm dw} (x) \cdot \bm{\tau} \right]
   \psi (x) =\epsilon \psi (x)~,
\end{equation}
where $\bm{\tau}$ is the vector of Pauli matrices and $\Delta$ is
the exchange splitting. For simplicity we have taken the system to
be one dimensional. We will further assume that the magnetization
texture is symmetric (in an obvious sense) around the domain-wall
position.

Asymptotically, the scattering state for an electron coming in
from the left reservoir with spin state $|\sigma\rangle$ is given
by
\begin{equation}
\label{eq:scatstatesleft}
  \psi^{1\sigma} (x) =
      \sum_{\sigma'} \left[ \delta_{\sigma  \sigma'} e^{i k_{\sigma'} x}  +
      \sqrt{\frac{k_\sigma}{k_{\sigma'}}} r_{11;\sigma'\sigma} e^{-i k_{\sigma'} x}\right]
      |\sigma'\rangle~,
\end{equation}
on the left, and
\begin{equation}
\label{eq:scatstatesright}
  \psi^{1\sigma} (x) =
      \sum_{\sigma'}  \sqrt{\frac{k_\sigma}{k_{-\sigma'}}} t_{12;\sigma'\sigma} e^{i k_{-\sigma'} x}
      |\sigma'\rangle~,
\end{equation}
on the right. The wave vectors are given by $k_\sigma = \sqrt{2m
(\epsilon+\sigma\Delta/2)/\hbar^2}$, with $\epsilon$ the electron
energy. (Note that the indices $\sigma,\sigma' \in
\{\uparrow,\downarrow\}$ refer to spin projections on the
$z$-axis, and that the respective number $\sigma$ takes on
$\{+1,-1\}$.)

We assume now that the amplitudes $r_{\sigma'\sigma}$ and
$t_{\sigma'\sigma}$ for a domain wall with zero chirality are
given, and will determine them numerically later. (See
Refs.~[\onlinecite{tatara2000}] and [\onlinecite{dugaev2003}] for
analytical expressions for these coefficients, valid for large and
small $k_F\lambda$, respectively.) All we need is that if we move
the domain wall to position $r_{\rm dw}$, the wave function $\psi
(x-r_{\rm dw})$ is a solution of the Schr\"odinger equation for an
electron moving in the displaced domain-wall texture. From this we
deduce the transmission and reflection amplitudes for arbitrary
domain-wall position. Furthermore, if the domain-wall precession
angle $\phi_0$, i.e., its chirality, becomes nonzero, the
magnetization texture changes to $\tilde{ \bm{\Omega}} (x) = (\sin
\theta_{\rm dw} (x) \cos \phi_0, \sin \theta_{\rm dw} (x) \sin
\phi_0, \cos \theta_{\rm dw} (x))$. The solution of the
Schr\"odinger equation for the wave function $\tilde \psi (x)$ of
electrons moving in this texture is given by
\begin{equation}
\label{eq:wavefctnonzerophi}
  \tilde \psi (x) = \left( \begin{array}{cc}
      e^{-i \phi_0/2} & 0 \\
      0 & e^{-i\phi_0/2}
    \end{array} \right) \psi (x)~,
\end{equation}
with $\psi (x)$ the solution of the Schrodinger equation in
Eq.~(\ref{eq:se}) for a zero-chirality domain wall. This is
sufficient to determine the chirality-dependence of the
transmission and reflection amplitudes.

Using these ingredients, we find for the reflection amplitude for
electrons off a domain wall at arbitrary position and with
arbitrary chirality that
\begin{eqnarray}
\label{eq:r11}
 {\bf r}_{11}  =\left( \begin{array}{ll}
      r_{\uparrow\uparrow} e^{2 i k_\uparrow r_{\rm dw}} & r_{\uparrow\downarrow} e^{i (k_\uparrow+k_\downarrow) r_{\rm dw}-i \phi_0} \\
      r_{\downarrow\uparrow} e^{i (k_\uparrow+k_\downarrow) r_{\rm dw}+i \phi_0}  &  r_{\downarrow\downarrow} e^{2 i k_\downarrow r_{\rm dw}}
    \end{array} \right),
\end{eqnarray}
and for the transmission amplitude
\begin{eqnarray}
\label{eq:t21}
 {\bf t}_{21}  =\left( \begin{array}{ll}
      t_{\uparrow\uparrow} e^{i (k_\uparrow-k_\downarrow) r_{\rm dw}} & t_{\uparrow\downarrow} e^{-i \phi_0} \\
      t_{\downarrow\uparrow} e^{i \phi_0}  &  t_{\downarrow\downarrow} e^{i (k_\downarrow-k_\uparrow) r_{\rm dw}}
    \end{array} \right).
\end{eqnarray}
The reflection and transmission amplitudes for electrons coming in
from the right reservoir, denoted by ${\bf r}_{22}$ and ${\bf
t}_{12}$, respectively, are determined in the same way and are
given by similar expressions. Because of the symmetries of the
domain wall we further have that $r_{11;\sigma\sigma} =
r_{22;-\sigma-\sigma}$, $t_{21;\sigma\sigma} =
t_{12;-\sigma-\sigma}$. Furthermore, $r_{11;\sigma\sigma'} =
-r_{22;\sigma'\sigma}$, and $t_{21;\sigma\sigma'} =
-t_{12;\sigma'\sigma}$, for $\sigma \neq \sigma'$. The final
results are given by
\begin{eqnarray}
\label{eq:r22}
 {\bf r}_{22}  =
  \left( \begin{array}{ll}
      r_{\downarrow\downarrow} e^{-2 i k_\uparrow r_{\rm dw}} & -r_{\downarrow\uparrow} e^{-i (k_\uparrow+k_\downarrow) r_{\rm dw}-i \phi_0} \\
      -r_{\uparrow\downarrow} e^{-i (k_\uparrow+k_\downarrow) r_{\rm dw}+i \phi_0}  &  r_{\downarrow\downarrow} e^{-2 i k_\downarrow r_{\rm dw}}
    \end{array} \right), \nonumber \\
\end{eqnarray}
and for the transmission amplitude
\begin{eqnarray}
\label{eq:t12}
 {\bf t}_{12}  =\left( \begin{array}{ll}
      t_{\downarrow\downarrow} e^{i (k_\uparrow-k_\downarrow) r_{\rm dw}} & -t_{\downarrow\uparrow} e^{-i \phi_0} \\
      -t_{\uparrow\downarrow} e^{i \phi_0}  &  t_{\uparrow\uparrow} e^{i (k_\downarrow-k_\uparrow) r_{\rm dw}}
    \end{array} \right).
\end{eqnarray}

\begin{center}
\begin{figure}
\epsfig{figure=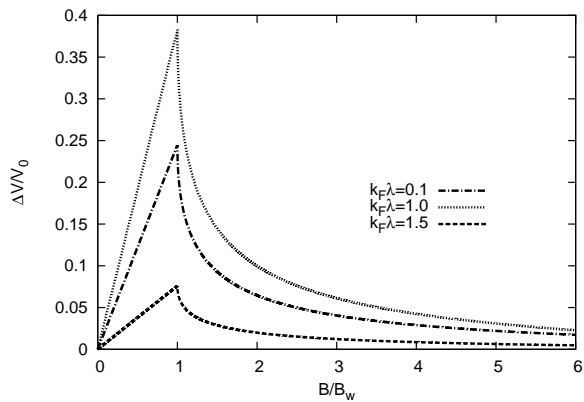, width=8cm}
 \caption{Voltage
  induced by a moving domain wall for Gilbert damping
  $\alpha_G=0.01$, for $\Delta/\epsilon_F=0.5$,
  and for various values of $k_F\lambda$. The voltage is normalized
  to $V_0 =g \mu_B B_w/|e|$.
  The magnetic field $B$ is in units of the Walker breakdown field $B_w$.
  For typical experiments we have that $V_0 \sim 0.5$ $\mu$Volt.}
 \label{fig:v_b_ballistic}
\end{figure}
\end{center}

Insertion of the results for the transmission and reflection
amplitudes into the expression for the current in
Eq.~(\ref{eq:currentviascatmatrix}), and identifying $X_1=r_{\rm
dw}$ and $X_2=\phi_0$, yields the result
\begin{widetext}
\begin{eqnarray}
\label{eq:resultballistic}
   I_1=-I_2&=&
 -\frac{|e|}{2\pi}
    \left[
      \sum_{\sigma}
     \left(2
     |r_{\sigma\sigma}|^2+|r_{\sigma-\sigma}|^2+|r_{-\sigma\sigma}|^2-|t_{\sigma\sigma}|^2+|t_{-\sigma-\sigma}|^2\right)
     k_\sigma
    \right] \frac{dr_{\rm dw} (t)}{dt} \nonumber \\
  &&  -\frac{|e|}{2\pi} \left( |r_{\downarrow\uparrow}|^2-|r_{\uparrow\downarrow}|^2-|t_{\downarrow\uparrow}|^2+|t_{\uparrow\downarrow}|^2\right)
  \frac{d\phi_0 (t)}{dt}~.
\end{eqnarray}
\end{widetext}
Fig.~\ref{fig:v_b_ballistic} gives the result for the voltage
\cite{footnote4} (that now includes the effect of the presence of
the domain wall on the conductance)
\begin{equation}
\label{eq:ballisticvoltage}
 \Delta V \equiv \frac{-I_1}{\frac{2\pi|e|^2}{\hbar}\left(\sum_\sigma  |t_{\sigma\sigma}|^2+|t_{\sigma-\sigma}|^2 \right)}~,
\end{equation}
for various values of $k_F \lambda$ and for the exchange splitting
$\Delta/\epsilon_F = 0.5$, for the case of a field-driven domain
wall which exhibits Walker breakdown. In this situation we have
that \cite{schryer1974}
\begin{eqnarray}
\label{eq:soleomvarparsaverage}
 \frac{d\phi_0 (t)}{dt} &=& \frac{1}{(1+\alpha_{G}^2)} \frac{g\mu_B B_w}{\hbar} {\rm Re}
 \left[ \sqrt{ \left( \frac{B}{B_w }\right)^2 - 1} \right]~; \nonumber \\
\frac{1}{\lambda}\frac{ d r_{\rm dw} (t)}{dt} &=& \frac{g \mu_B
B}{\alpha_{G} \hbar} -  \frac{1}{\alpha_G} \frac{d \phi_0
(t)}{dt}~.
\end{eqnarray} The various transmission and reflection amplitudes are determined
by numerically solving the Schr\"odinger equation in
Eq.~(\ref{eq:se}), for the specific domain-wall profile
\cite{tatara2004}
\begin{equation}
 \theta_{\rm dw} (x) = 2 \tan^{-1} \left( e^{\frac{x}{\lambda}}\right)~.
\end{equation}
This texture corresponds to a domain wall that interpolates
between a domain with magnetization point in the $+z$-direction,
to a domain with magnetization in the opposite direction (see
Fig.~\ref{fig:dw}).

The induced voltage in Fig.~\ref{fig:v_b_ballistic} is for fields
below Walker breakdown qualitatively similar to that in
Fig.~\ref{fig:v_b} for the diffusive case. The voltage is largest
when $k_F \lambda \simeq 1$. This is to be expected, as the
domain-wall resistance is also largest for that situation,
\cite{tatara2000,dugaev2003} and therefore non-adiabatic effects
are also large. We also find that the voltage becomes larger upon
increasing the ratio $\Delta/\epsilon_F$.

The ballistic model presented here underestimates the voltage for
fields above Walker breakdown. This is most easily understood by
taking the adiabatic limit $k_F\lambda \to \infty$. In that case
all reflection coefficients $r_{\sigma\sigma'} \to 0$, the
diagonal transmission coefficients $t_{\sigma\sigma} \to 0$, and
the off-diagonal transmission coefficients $t_{\sigma\sigma'} \to
1$. In this limit we find in first instance that, according to
Eq.~(\ref{eq:resultballistic}), the induced current and voltage
become zero. This is a result of the simplicity of the model
presented here and could be repaired by incorporating more
transverse channels. (See Refs.~[\onlinecite{waintal2004}] and
[\onlinecite{mazin1999}] for a discussion of subtleties in
describing spin-polarized transport with the Landauer-B\"uttiker
formalism.) To establish connection of the result of our simple
ballistic model with the result found in the diffusive limit, it
is easiest to put the spin polarization of the current in by hand.
This is done by weighing the spin-up and down electrons coming
from the left reservoir differently. In the adiabatic limit this
amounts to taking $|t_{\downarrow\uparrow}|^2=(1+{\mathcal P})/2$,
$|t_{\uparrow\downarrow}|^2=(1-{\mathcal P})/2$, and putting all
other coefficients equal to zero. With this modification the
result in
Eqs.~(\ref{eq:resultballistic})~and~(\ref{eq:ballisticvoltage})
reduces to that in Eq.~(\ref{eq:resultcurrentafteransatz2}) with
$\beta_{\rm sf}=\beta_{\rm na}=0$.

We end by remarking that the ballistic model presented in this
section applies when the phase-coherence length of the electrons
is at least of the order of the domain-wall width. Second, because
we have taken only one (spin-resolved) transport channel our
calculation applies only to very tightly confined structures such
as ferromagnetic nanocontacts. \cite{garcia1999} (Alhough it may
well be that the model for the domain-wall motion in
Eq.~(\ref{eq:soleomvarparsaverage}) needs to be refined to apply
to such situations.) Nonetheless, the method presented in this
section should, when modified to apply to more realistic
geometries, be a convenient starting point for addressing the
ballistic case should experiments approach this limit in the near
future.

\section{Discussion and conclusions} \label{sec:disc} We have
discussed the effects of non-adiabaticity on the voltage induced
by a field-driven domain wall, and considered both the diffusive
and ballistic regimes of electronic transport. In particular, we
have shown that incorporating effects of non-adiabaticity of the
wall on the induced voltage is done by the same renormalization
that incorporates non-adiabaticity in the description of
current-induced propagation.

Future work will include incorporating also the effects of
anisotropic magnetoresistance on the induced voltage, and more
sophisticated models of domain-wall motion, such as vortex walls.
(We mention however that the rigid-domain-wall model discussed in
this paper is known to give sensible results below Walker
breakdown. \cite{beach2008}) Another interesting subject for study
is the effect of spin relaxation and non-adiabaticity on the
motive forces in mesoscopic rings,\cite{stern1992} where these
effects may actually be absent.

We hope that the present work, as well as previous theoretical
work,
\cite{berger1986,stern1992,barnes2007,duine2008,tserkovnyak2008,saslow2007,ohe2007,xiao2008,stamenova2008}
will motivate experimental progress in the direction of observing
motive forces and voltages by time-dependent magnetization
textures. One interesting aspect would be the possibility of
determinating of the degree of non-adiabaticity and effects of
spin relaxation directly from such experiments.\cite{duine2008}

It is a great pleasure to thank Stewart Barnes, Gerrit Bauer,
Mathias Kl\"aui, Hiroshi Kohno, Ties Lucassen, Allan MacDonald,
Sadamichi Maekawa, Wayne Saslow, Mark Stiles, Henk Stoof, Henk
Swagten, Yaroslav Tserkovnyak for discussions. I would also like
to gratefully acknowledge Henk Stoof for carefully reading the
manuscript. This work was supported by the Nederlandse Organisatie
voor Wetenschappelijke Onderzoek (NWO).


\begin{thebibliography}{99}
\bibitem{footnote1} Here, ``non-adiabatic" refers to terms beyond linear
response in spatial gradients of the magnetization direction,
considering, however, only first order in temporal derivatives.
\bibitem{footnote3} Note $I_\nu$ is a
current with units $[I_\nu]=$ C/s, whereas $j$ is strictly
speaking a current density with $[j]=$ C/m$^2$s.
\end{thebibliography}
\end{document}